\documentclass[twocolumn,showpacs,preprintnumbers,amsmath,amssymb]{revtex4}
\usepackage{epsfig}
\usepackage{amsmath}
\usepackage{bm}% bold math

\newcommand{\be}{\begin{equation}}
\newcommand{\ee}{\end{equation}}
\newcommand{\ket}[1]{\left| #1 \right\rangle}
\newcommand{\sub}[1]{_{\mbox{\scriptsize #1}}}
\renewcommand{\Re}{\mbox{Re}\:}

\begin{document}

\title{High-fidelity manipulation of a Bose-Einstein condensate using Bragg 
interactions}

\author{K. J. Hughes, B. Deissler, J. H. T. Burke and C.A. Sackett}
\affiliation{Physics Department, University of Virginia, Charlottesville, VA 22904}
\email{sackett@virginia.edu}
\date{\today}

\begin{abstract}
The use of off-resonant standing light waves to manipulate ultracold 
atoms is investigated.  Previous work has illustrated that
optical pulses can provide efficient beam-splitting and reflection
operations for atomic wave packets.  The performance of these operations
is characterized experimentally using Bose-Einstein condensates
confined in a weak magnetic trap. 
Fidelities of 0.99 for beam splitting and 0.98 for 
reflection are observed, and splitting operations of up to third order
are achieved.
The dependence of the operations on light intensity and atomic velocity  
is measured and found to agree well with theoretical estimates.
\end{abstract}


\maketitle

Optical control of atomic motion has developed into a fertile
field over the past few decades.  One important tool is the 
off-resonant standing wave laser beam \cite{Metcalf99,Meystre01}.    
Atoms in such a beam experience a spatially oscillating potential that 
can act as
a diffraction grating for the atomic wave function in the same way that
a conventional grating acts for a light wave.  
This effect is referred to as Bragg diffraction.
Bragg diffraction was originally demonstrated by deflecting
thermal atomic beams \cite{Martin88,Giltner95},
and the technique has seen much use in atom-beam interferometers
\cite{Berman97}.
More recently,
similar effects have been achieved using nearly stationary atoms produced
using laser cooling or Bose-Einstein 
condensation \cite{Kunze96,Kozuma99}.  This too has been applied 
to atom interferometers 
\cite{Torii00,Wang05,Wu05,Garcia06,Horikoshi06}
and other experiments \cite{Hagley99b,Inouye99,Kinoshita06}.

Although several studies of Bragg diffraction were performed using
atomic beams \cite{Martin88,Giltner95,Berman97}, the large spread in
atomic velocity complicated comparisons with theory and also
limited the fidelity with which operations such as beam-splitting
and reflection could be applied.  Accurate operations are 
advantageous for most applications.  
For instance, in
nearly all interferometer schemes, imperfect operations will reduce
signal visibility through the loss of atoms and the presence of 
non-interfering atoms in the measured output states.
More specifically, for interferometers such as
in \cite{Wang05,Garcia06}, a stationary condensate is split into
two packets moving apart.  After some time the packets are brought
back together and returned to zero velocity.  Any residual atoms left
at zero velocity after the splitting operation would interfere with
the final recombination and introduce phase errors.  

With ultracold atoms, quite accurate Bragg
manipulations can be achieved.  This has been observed
in experiments for some time, but a quantitative study has not been performed.  
In this paper, we investigate the dependence of the operations on 
two key parameters, the
laser intensity and residual atomic velocity, and find good agreement
with theoretical expectations.   We also report a 
level of fidelity that has not, to our knowledge, been previously seen.  
This demonstrates the possibility to control atomic motion
with a degree of precision
comparable to that previously achievable only with internal state transitions.

We focus attention on the beam-splitting operation first implemented
in \cite{Wang05} and the reflection operation demonstrated in
\cite{Garcia06}.  
A general understanding of these manipulations can be obtained
by considering
the three states $\ket{0}, \ket{+v_0}$ and $\ket{-v_0}$, where the state
labels give the atomic velocity and $v_0 = 2\hbar k/M$ 
for atomic mass $M$ and light wavenumber $k$.  
As will be detailed below,
the optical standing wave potential can
be expressed as $\hbar\beta\cos 2ky$.
This couples the
state $\ket{0}$ to the states $\ket{\pm v_0}$ via matrix elements
$\hbar\beta/2$, driving the beam-splitting transition 
\begin{equation}
\ket{0} \leftrightarrow \ket{+} \equiv \frac{1}{\sqrt{2}}\left(
\ket{v_0} + \ket{-v_0}\right).
\end{equation}
However, since the $\ket{0}$ and $\ket{+}$ states have energies that
differ by $mv_0^2/2$ the transition is not resonant and cannot be made
with perfect efficiency using a single pulse.  An appropriate pulse can,
however, create the superposition state
$(\ket{0} + \ket{+})/\sqrt{2}$.  Due to the energy 
difference, the phase of this superposition changes as the state
evolves freely in time, eventually becoming $(\ket{0} - \ket{+})/\sqrt{2}$.  
If a second identical pulse is then applied, 
the state evolves to the desired $\ket{+}$.
This analysis follows that of Wu {\em et al.} \cite{Wu05b}.

The reflection operation $\ket{+v_0} \leftrightarrow \ket{-v_0}$ couples
two states of equal energy, so multiple pulses are not required.  However,
the coupling between them is only second order, making the transition
rather slow.  Efficient operation can be obtained by noting that the state
$\ket{+v_0}$ can be expressed as $(\ket{+}+\ket{-})/\sqrt{2}$, where
the antisymmetric state $\ket{-} = (\ket{+v_0} - \ket{-v_0})/\sqrt{2}$
has no coupling to the $\ket{0}$ state.  The $\ket{-}$ state does, 
however, acquire an
energy shift while the Bragg beam is on, making its phase evolve in time.
For the appropriate pulse intensity and duration, the $\ket{+}$ branch of the
wave function makes two full Rabi oscillations between $\ket{0}$ and 
$\ket{+}$, ending up back in $\ket{+}$ as it started.  At the same 
time, the $\ket{-}$ branch acquires a $\pi$ phase shift, making the
total state $(\ket{+}-\ket{-})/\sqrt{2} = \ket{-v_0}$ and achieving the
desired reflection.  

To study these operations, we applied them to Bose-Einstein
condensates consisting
of about $10^4$ $^{87}$Rb atoms held in a magnetic guide.  The 
apparatus has been described previously \cite{Reeves05}.  
One feature of the guide is its relatively weak confinement:
for the experiments described here, the harmonic oscillation
frequencies were $\omega_x = 2\pi\times 7.4$~Hz, $\omega_y = 2\pi\times
0.8$~Hz, and $\omega_z = 2\pi\times 4.3$~Hz, with the Bragg laser beam
parallel to the $y$-axis.  As a result, the atomic densities are
relatively low and interaction effects are negligible.

Another relevant aspect of the apparatus is the procedure by which atoms
are loaded into the guide.  The condensates are produced in a much 
tighter trap suitable for evaporative cooling.  The tight trap and the
guide are coincident, and the guide is loaded by slowly ramping up the
guide field and then ramping down the tight trap field. 
During the final ramp, the trap frequencies pass through a
value of 60 Hz, at which point ambient fields
with a few mG amplitude can excite center-of-mass
oscillation of atoms.  Velocities of up to 0.5 mm/s were observed
as a result, which is large enough
to affect the splitting and reflection operations.  
Removing nearby noise sources reduced the velocity amplitude to
about 0.2 mm/s.  This was further controlled by
synchronizing the loading
process with the power line frequency and adjusting the start of
the experiment to occur at at turning point of the atomic motion where 
the velocity was near zero.

Alternatively, we could use the loading process to controllably
impart a velocity to the atoms.  In this case, 
the centers of the tight trap and the guide were deliberately offset,
and we suddenly turned off the tight trap current before it reached
zero.  This caused the atoms to oscillate with a large amplitude.
By starting the experiment at different points in the cycle,
various atomic velocities were sampled and the effect of 
velocity on the Bragg operations could be investigated.
Since the Bragg operations require less than 1 ms to perform, the
atomic velocity was essentially constant during the experiment.

The Bragg standing wave is produced by a home-built diode laser.  The
laser is tuned 12.8 GHz blue of the 5S$_{1/2}$ to 5P$_{3/2}$
laser cooling transition, at a wavelength of 780.220~nm.  
The frequency is stabilized using a
cavity-transfer lock \cite{Burke05}
referenced to a resonant laser locked via saturated absorption.  
An acousto-optic modulator (AOM) is used to switch the beam on and
off, and also to adjust the beam intensity.  The output of the AOM is
coupled into a single-mode optical fiber that provides spatial filtering
and pointing stabilization.  The fiber output has a power of up to
12 mW and an approximately 
Gaussian profile with beam waist of 0.7~mm.  The standing wave
is generated by passing the beam through the vacuum cell and retro-reflecting 
it using an external mirror.
The cell was constructed with vacuum windows that were anti-reflection
coated on both sides.  We found this to be critical: The cell used in
Ref.~\cite{Garcia06} had uncoated Pyrex windows, and the multiple reflections 
produced a speckle pattern in the reflected beam with intensity variations
of over 50\%.  This variation made it difficult to achieve consistent
results.  

The experiments were performed by applying one or more pulses
of the Bragg beam and then allowing the atoms to evolve freely for 30 ms so 
that packets with different velocities would separate in space.
The atoms were observed by passing a resonant probe beam through 
them into a camera, with the resulting 
images showing the positions of the packets and the
number of atoms in each.  The images were analyzed by fitting the
packets to Gaussian profiles.  For faint packets, the 
widths of the profiles were fixed to match those 
observed for packets containing many atoms, typically 10 $\mu$m
and 70 $\mu$m in the $x$ and $y$ directions, respectively.

In general, several packets are observed in a given image.  The
packets are labeled according to their velocity index $n$, 
defined by $v = v_i + nv_0$
where $v$ is the observed velocity and $v_i$ the initial velocity.
The fraction of atoms in each packet is denoted $N_n$.  The accuracy
of the operation is quantified using the fidelity, defined as
\be
F = \left|\langle\psi_0 | \psi\rangle\right|^2
\ee
for desired state $\psi_0$ and observed state $\psi$.  For instance,
an ideal splitting operation produces the state $\ket{+}$, so if the
actual state is 
\be
\ket{\psi} = \sum_n c_n\ket{v_i + n v_0} 
\ee
then the fidelity would be
\be
F\sub{split} = \frac{1}{2}\left|c_{+1} + c_{-1}\right|^2 = 
\frac{1}{2}\left(N_{+1} + N_{-1} + 2\Re{c_{+1}^*c_{-1}}\right).
\ee
In general this depends on the phase difference between $c_{+1}$ and
$c_{-1}$, which of course cannot be determined by simple imaging.  
Moreover, interferometer operation does not depend
on this phase either as long as it is constant, 
since any phase difference imposed by the splitting
operation can be absorbed into the definitions of the states $\ket{\pm v_0}$ 
without changing the ultimate results.  For this reason, we ignore
any phase effects and estimate the fidelity as
\be
F\sub{split} = \frac{1}{2}\left(N_{+1} + N_{-1} + 2\sqrt{N_{+1} N_{-1}}\right).
\ee

We observe the most accurate splitting operation when using 
zero-velocity atoms and a Bragg beam power of 0.34 mW. 
Figure 1 shows the results obtained 
as the power and velocity were varied around these values.
Note that the velocity
width observed here corresponds to a temperature of 60 nK,  
which illustrates the benefit of performing such experiments using
condensates.  Each data point shown represents a single measurement.
Repeated measurements under the same conditions showed varations of about
about 0.01 in $N_0/N$. 
At the optimal parameters,
the best images yielded fidelities of about 0.995, with repeated measurements
consistently above 0.99.  Image noise limits the accuracy of our analysis 
to about this level, but we also typically observe longer-term variations 
that we attribute to experimental variations in the atomic velocity, beam
alignment, and laser power.  Nonetheless it was possible to maintain
split fidelities near 0.99 over a period of several hours.

\begin{figure}
\epsfig{file=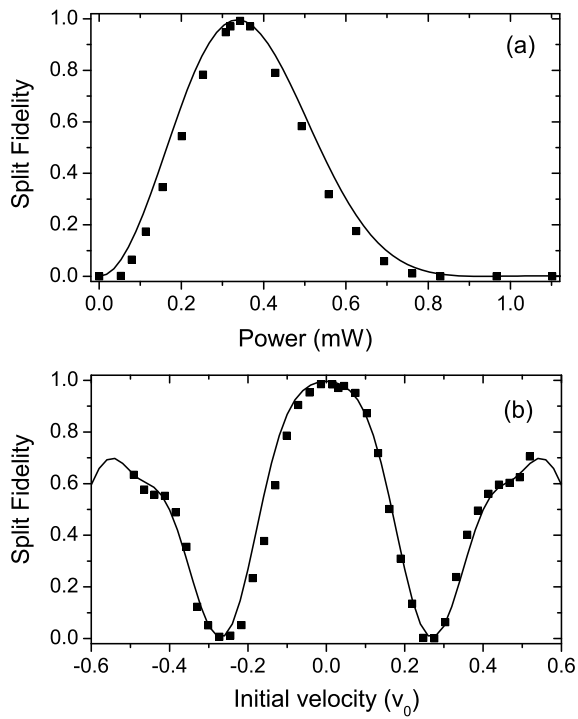,width=3in}
\caption{Fidelity of splitting operation.  Data points show
the experimentally measured fidelity as a function of (a) the Bragg
laser power and (b) initial atomic velocity.  
The solid curves are
theoretical calculations and include a calibration adjustment of 1.19
to relate the standing wave intensity to the measured power.  The pulse 
sequence consisted of a pulse with duration 24~$\mu$s, a 33~$\mu$s delay, and
a second pulse identical to the first.}
\end{figure}

Similar results for the reflection operation are shown in Fig.~2.
Here the fidelity is given by $N_{+2}$, since the initial velocity is
close to $-v_0$.  The optimal  fidelity was about 0.94.

The theoretical curves shown in Figs. 1 and 2 are calculated by 
solving the Schr\"odinger equation in the presence of the optical
potential,
\be
i\frac{d\psi}{dt} = \left[-\frac{\hbar}{2M} \frac{\partial^2\psi}{\partial y^2}
+\beta \cos(2 k y) \psi\right],
\ee
using the Bloch expansion
\be
\psi(y,t) = \sum_n c_n(t) e^{i(2nk + \kappa)y},
\ee
where $\kappa = Mv_i/\hbar$ \cite{Wu05b}.
The coefficients $c_n$ then satisfy
\be
\label{raman-nath}
i\frac{dc_n}{dt} = \frac{\hbar}{2M}(2nk+\kappa)^2c_n + \frac{\beta}{2}
(c_{n-1} + c_{n+1}).
\ee
We solved equations (\ref{raman-nath}) numerically, including indices $n$
up to $\pm 4$.  (Negligible population was observed in the $c_{\pm 4}$
states.)  We applied the initial condition $c_0 = 1$ and calculated
the fidelity as described above.  The results show good agreement with
the experimental data, but to achieve this it was necessary to 
calibrate the standing wave amplitude $\beta$.  

\begin{figure}
\epsfig{file=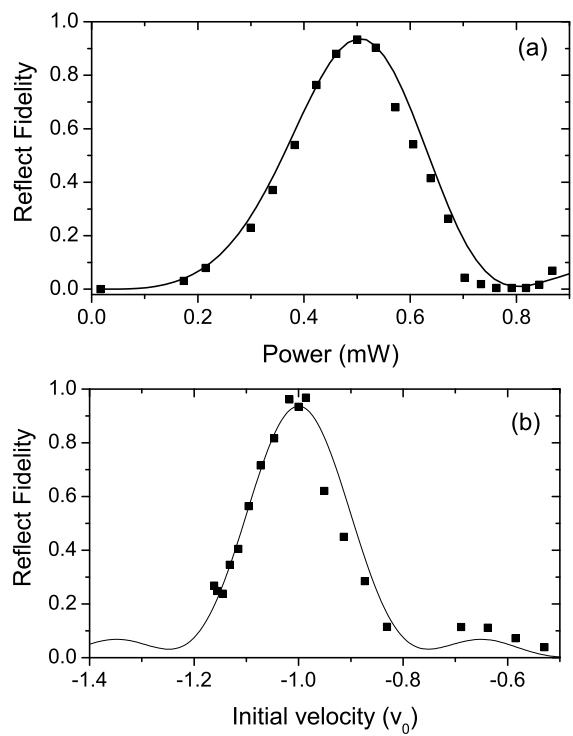,width=3in}
\caption{Fidelity of reflection operation.  Data points show the measured
values and curves are a theoretical calculation.  The operation consists of 
a single pulse with duration 76~$\mu$s.}
\end{figure}

To calculate $\beta$, we treat the atoms as two-level systems,
in which the $5S{1/2} F =2$ ground hyperfine state is coupled to the
$5P_{3/2}$ excited state.  
We neglect the excited state hyperfine splitting, since it is small
compared to the laser detuning $\Delta$.  The optical potential
is then given by \cite{Metcalf99}
\be
V(y) = \frac{\hbar \Gamma^2}{8\Delta} \frac{I(y)}{I\sub{sat}}
\ee
where $\Gamma = 3.8\times10^7$~s$^{-1}$ 
is the excited state linewidth, $I$ is laser intensity,
and $I\sub{sat} = 2.5$~mW/cm$^2$ is the saturation intensity
for linearly polarized light.
For an incident beam with power $P$ and Gaussian waist $w$, the 
standing wave intensity is given by
\be
I(y) = \frac{8P}{\pi w^2}\cos^2(ky) = \mbox{const} + \frac{4P}{\pi w^2}
\cos 2ky.
\ee
The constant term has no effect, so comparison with the form
$V = \hbar\beta\cos(2ky)$ gives
\begin{equation}
\label{eq-beta}
\beta = \frac{\Gamma^2 P}{2\pi\Delta I_s w^2} \approx 10\hbar\omega_r 
\times \frac{P}{\mbox{mW}} 
\end{equation}
where the recoil frequency $\omega_r$ is defined as $\hbar k^2/2M
\approx 2.36\times 10^4$~s$^{-1}$.  The optimum theoretical fidelity
for the split operation
occurs at $\beta = 2.8\omega_r$, or $P = 0.28$ mW,  about
20\% different from the observed peak.  We attribute this discrepancy 
to experimental errors in the intensity calibration.  In particular,
intensity variations of this order can result from  
small misalignments of the beam on the atoms or
deviations of the beam profile from a Gaussian.
We therefore include
a calibration factor in the theory curve, and adjust it to match the
measured peak.  
The ratio between the powers for the splitting and reflection 
operations was about 0.65, in good agreement with theoretical 
expectation $\beta\sub{reflect} = 4.4\omega_r$.

\begin{figure}
\epsfig{file=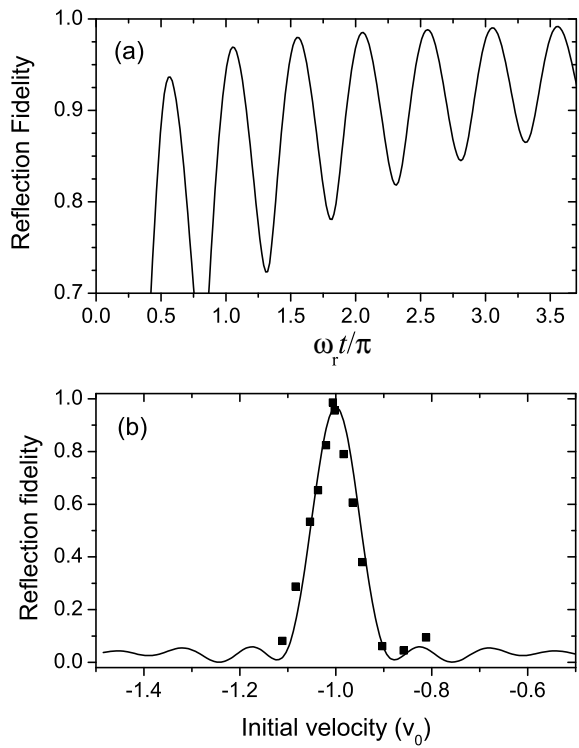}
\caption{(a) Optimized reflection fidelity vs. pulse duration.  For
each pulse duration $t$, the standing wave amplitude $\beta$ was adjusted
to provide the best reflection fidelity for atoms traveling at
$v_i = v_0$.  The results of Fig.~2 correspond to the peak near
$\omega_r t = \pi/2$.  (b) Experimental results for a longer pulse.
Here $\omega_r t = 3.3$ and the beam power was 0.37~mW.}
\end{figure}

It is noteworthy that the optimum reflection fidelity is 
lower than that achievable for the splitting operation.  One solution
is to use a longer pulse with lower intensity \cite{Giltner95}.  
In the limit of a very
long, very weak pulse, perfect reflection should be possible.  
We show how this regime can be approached in Fig.~3(a).  This curve
was calculated by varying the pulse length and at each point numerically
optimizing the value of $\beta$.  It is observed that the reflection
fidelity does indeed improve, though not monotonically.  This behavior
can be understood with reference to the three-level model 
discussed previously.  If the problem is solved in detail, it is found
that the initial state
$\ket{+v_0} = (\ket{-}+\ket{+})/\sqrt{2}$ evolves to 
\be
\begin{split}
\ket{\psi} = \frac{e^{-2i\omega_r t}}{\sqrt{2}} \Big[ & 
e^{-2i\omega_r t}\ket{-} 
-i\frac{\beta\sqrt{2}}{X}\sin\frac{Xt}{2}\ket{0}
 \\
{} &+\left(\cos\frac{Xt}{2}-i\frac{4\omega_r}{X}\sin\frac{Xt}{2}\right)\ket{+}
\Big]
\end{split}
\ee
where $X = \left(2\beta^2+16\omega_r^2\right)^{1/2}$.  The desired state
$\ket{-v_0} = (\ket{+}-\ket{-})/\sqrt{2}$ is achieved when 
$2\omega_r t = n\pi$ and $Xt/2 = (n+1)\pi$ for integer $n > 0$.  The
peaks observed in Fig.~3 indeed occur at $\omega_r t\approx n\pi/2$.
Solving for the required intensity yields
$\beta = 2\omega_r (4n+2)^{1/2}/n$.
The reflection operation is more accurate for large $n$ because 
coupling to higher velocity states is reduced as $\beta$ decreases.

We experimentally observed the second maximum, at 
pulse time $t = 140~\mu$s and $\beta = 3.0 \omega_r$.  The results
are shown in Fig.~3(b), where a peak fidelity of about 0.98 is observed.  The
longer pulse duration, however, results in greater sensitivity to the atom
velocity.  In situations where the velocity cannot be perfectly controlled,
the shorter pulse may be preferable.

Reference \cite{Wu05b} 
also points out that the double-pulse splitting sequence 
can be used efficiently for higher order operations, driving
for instance $\ket{0} \rightarrow (\ket{+2v_0}+\ket{-2v_0})/\sqrt{2}$.
We implemented these operations as well.  For the order-2 split
above, we observed a peak fidelity of about 0.92, while the order-3
split produced atoms with velocity $\pm 3v_0$ at a fidelity of 
0.8.  Theory predicts fidelities of 0.99 and 0.97 respectively.  
For at least the third-order result, the relatively 
poor experimental performance can be attributed to the high
sensitivity of the operation to the pulse intensity
and duration.  Our timing resolution is limited by
the 100-ns switching time of the AOM, and we observed the order-3 pulse
to be sensitive on that time scale.  We did not implement an order-4
split because it requires greater intensity than we had available.

In summary, we have investigated the optical 
manipulation of atomic wave packets.  We demonstrated
splitting and reflecting with high fidelity and good agreement
with theoretical expectations.  We hope that these results will
prove useful for the development of atom interferometers and other
applications of atom optics.

This work was sponsored by the Defense Advanced Research Projects
Agency (award No. 51925-PH-DRP)
and by the National Science Foundation (award No. PHY-0244871).
We thank Eun Oh and David Weiss for stimulating conversations.

%\bibliographystyle{apsrev}
%\bibliography{sackett}

\begin{thebibliography}{18}
\expandafter\ifx\csname natexlab\endcsname\relax\def\natexlab#1{#1}\fi
\expandafter\ifx\csname bibnamefont\endcsname\relax
  \def\bibnamefont#1{#1}\fi
\expandafter\ifx\csname bibfnamefont\endcsname\relax
  \def\bibfnamefont#1{#1}\fi
\expandafter\ifx\csname citenamefont\endcsname\relax
  \def\citenamefont#1{#1}\fi
\expandafter\ifx\csname url\endcsname\relax
  \def\url#1{\texttt{#1}}\fi
\expandafter\ifx\csname urlprefix\endcsname\relax\def\urlprefix{URL }\fi
\providecommand{\bibinfo}[2]{#2}
\providecommand{\eprint}[2][]{\url{#2}}

\bibitem[{\citenamefont{Metcalf and van~der Straten}(1999)}]{Metcalf99}
\bibinfo{author}{\bibfnamefont{H.~J.} \bibnamefont{Metcalf}} \bibnamefont{and}
  \bibinfo{author}{\bibfnamefont{P.}~\bibnamefont{van~der Straten}},
  \emph{\bibinfo{title}{Laser Cooling and Trapping}}
  (\bibinfo{publisher}{Springer}, \bibinfo{address}{New York},
  \bibinfo{year}{1999}).

\bibitem[{\citenamefont{Meystre}(2001)}]{Meystre01}
\bibinfo{author}{\bibfnamefont{P.}~\bibnamefont{Meystre}},
  \emph{\bibinfo{title}{Atom Optics}} (\bibinfo{publisher}{Springer},
  \bibinfo{address}{New York}, \bibinfo{year}{2001}).

\bibitem[{\citenamefont{Martin et~al.}(1988)\citenamefont{Martin, Oldaker,
  Miklich, and Pritchard}}]{Martin88}
\bibinfo{author}{\bibfnamefont{P.~J.} \bibnamefont{Martin}},
  \bibinfo{author}{\bibfnamefont{B.~G.} \bibnamefont{Oldaker}},
  \bibinfo{author}{\bibfnamefont{A.~H.} \bibnamefont{Miklich}},
  \bibnamefont{and} \bibinfo{author}{\bibfnamefont{D.~E.}
  \bibnamefont{Pritchard}}, \bibinfo{journal}{Phys. Rev. Lett.}
  \textbf{\bibinfo{volume}{60}}, \bibinfo{pages}{515} (\bibinfo{year}{1988}).

\bibitem[{\citenamefont{Giltner et~al.}(1995)\citenamefont{Giltner, McGowan,
  and Lee}}]{Giltner95}
\bibinfo{author}{\bibfnamefont{D.~M.} \bibnamefont{Giltner}},
  \bibinfo{author}{\bibfnamefont{R.~W.} \bibnamefont{McGowan}},
  \bibnamefont{and} \bibinfo{author}{\bibfnamefont{S.~A.} \bibnamefont{Lee}},
  \bibinfo{journal}{Phys. Rev. A} \textbf{\bibinfo{volume}{52}},
  \bibinfo{pages}{3966} (\bibinfo{year}{1995}).

\bibitem[{\citenamefont{Berman}(1997)}]{Berman97}
\bibinfo{editor}{\bibfnamefont{P.~R.} \bibnamefont{Berman}}, ed.,
  \emph{\bibinfo{title}{Atom Interferometry}} (\bibinfo{publisher}{Academic
  Press}, \bibinfo{address}{San Diego}, \bibinfo{year}{1997}).

\bibitem[{\citenamefont{Kunze et~al.}(1996)\citenamefont{Kunze, D{\"u}rr, and
  Rempe}}]{Kunze96}
\bibinfo{author}{\bibfnamefont{S.}~\bibnamefont{Kunze}},
  \bibinfo{author}{\bibfnamefont{S.}~\bibnamefont{D{\"u}rr}}, \bibnamefont{and}
  \bibinfo{author}{\bibfnamefont{G.}~\bibnamefont{Rempe}},
  \bibinfo{journal}{Europhys. Lett.} \textbf{\bibinfo{volume}{34}},
  \bibinfo{pages}{343} (\bibinfo{year}{1996}).

\bibitem[{\citenamefont{Kozuma et~al.}(1999)\citenamefont{Kozuma, Deng, Hagley,
  Wen, Lutwak, Helmerson, Rolston, and Phillips}}]{Kozuma99}
\bibinfo{author}{\bibfnamefont{M.}~\bibnamefont{Kozuma}},
  \bibinfo{author}{\bibfnamefont{L.}~\bibnamefont{Deng}},
  \bibinfo{author}{\bibfnamefont{E.~W.} \bibnamefont{Hagley}},
  \bibinfo{author}{\bibfnamefont{J.}~\bibnamefont{Wen}},
  \bibinfo{author}{\bibfnamefont{R.}~\bibnamefont{Lutwak}},
  \bibinfo{author}{\bibfnamefont{K.}~\bibnamefont{Helmerson}},
  \bibinfo{author}{\bibfnamefont{S.~L.} \bibnamefont{Rolston}},
  \bibnamefont{and} \bibinfo{author}{\bibfnamefont{W.~D.}
  \bibnamefont{Phillips}}, \bibinfo{journal}{Phys. Rev. Lett.}
  \textbf{\bibinfo{volume}{82}}, \bibinfo{pages}{871} (\bibinfo{year}{1999}).

\bibitem[{\citenamefont{Torii et~al.}(2000)\citenamefont{Torii, Suzuki, Kozuma,
  Sugiura, Kuga, Deng, and Hagley}}]{Torii00}
\bibinfo{author}{\bibfnamefont{Y.}~\bibnamefont{Torii}},
  \bibinfo{author}{\bibfnamefont{Y.}~\bibnamefont{Suzuki}},
  \bibinfo{author}{\bibfnamefont{M.}~\bibnamefont{Kozuma}},
  \bibinfo{author}{\bibfnamefont{T.}~\bibnamefont{Sugiura}},
  \bibinfo{author}{\bibfnamefont{T.}~\bibnamefont{Kuga}},
  \bibinfo{author}{\bibfnamefont{L.}~\bibnamefont{Deng}}, \bibnamefont{and}
  \bibinfo{author}{\bibfnamefont{E.~W.} \bibnamefont{Hagley}},
  \bibinfo{journal}{Phys. Rev. A} \textbf{\bibinfo{volume}{61}},
  \bibinfo{pages}{041602(R)} (\bibinfo{year}{2000}).

\bibitem[{\citenamefont{Wang et~al.}(2005)\citenamefont{Wang, Anderson, Bright,
  Cornell, Diot, Kishimoto, Prentiss, Saravanan, Segal, and Wu}}]{Wang05}
\bibinfo{author}{\bibfnamefont{Y.~J.} \bibnamefont{Wang}},
  \bibinfo{author}{\bibfnamefont{D.~Z.} \bibnamefont{Anderson}},
  \bibinfo{author}{\bibfnamefont{V.~M.} \bibnamefont{Bright}},
  \bibinfo{author}{\bibfnamefont{E.~A.} \bibnamefont{Cornell}},
  \bibinfo{author}{\bibfnamefont{Q.}~\bibnamefont{Diot}},
  \bibinfo{author}{\bibfnamefont{T.}~\bibnamefont{Kishimoto}},
  \bibinfo{author}{\bibfnamefont{M.}~\bibnamefont{Prentiss}},
  \bibinfo{author}{\bibfnamefont{R.~A.} \bibnamefont{Saravanan}},
  \bibinfo{author}{\bibfnamefont{S.~R.} \bibnamefont{Segal}}, \bibnamefont{and}
  \bibinfo{author}{\bibfnamefont{S.}~\bibnamefont{Wu}}, \bibinfo{journal}{Phys.
  Rev. Lett.} \textbf{\bibinfo{volume}{94}}, \bibinfo{pages}{090405}
  (\bibinfo{year}{2005}).

\bibitem[{\citenamefont{Wu et~al.}(2005{\natexlab{a}})\citenamefont{Wu, Su, and
  Prentiss}}]{Wu05}
\bibinfo{author}{\bibfnamefont{S.}~\bibnamefont{Wu}},
  \bibinfo{author}{\bibfnamefont{E.~J.} \bibnamefont{Su}}, \bibnamefont{and}
  \bibinfo{author}{\bibfnamefont{M.}~\bibnamefont{Prentiss}},
  \bibinfo{journal}{Euro. Phys. J. D} \textbf{\bibinfo{volume}{35}},
  \bibinfo{pages}{111} (\bibinfo{year}{2005}{\natexlab{a}}).

\bibitem[{\citenamefont{Garcia et~al.}(2006)\citenamefont{Garcia, Deissler,
  Hughes, Reeves, and Sackett}}]{Garcia06}
\bibinfo{author}{\bibfnamefont{O.}~\bibnamefont{Garcia}},
  \bibinfo{author}{\bibfnamefont{B.}~\bibnamefont{Deissler}},
  \bibinfo{author}{\bibfnamefont{K.~J.} \bibnamefont{Hughes}},
  \bibinfo{author}{\bibfnamefont{J.~M.} \bibnamefont{Reeves}},
  \bibnamefont{and} \bibinfo{author}{\bibfnamefont{C.~A.}
  \bibnamefont{Sackett}}, \bibinfo{journal}{Phys. Rev. A}
  \textbf{\bibinfo{volume}{74}}, \bibinfo{pages}{031601(R)}
  (\bibinfo{year}{2006}).

\bibitem[{\citenamefont{Horikoshi and Nakagawa}(2006)}]{Horikoshi06}
\bibinfo{author}{\bibfnamefont{M.}~\bibnamefont{Horikoshi}} \bibnamefont{and}
  \bibinfo{author}{\bibfnamefont{K.}~\bibnamefont{Nakagawa}},
  \bibinfo{journal}{Phys. Rev. A} \textbf{\bibinfo{volume}{74}},
  \bibinfo{pages}{031602(R)} (\bibinfo{year}{2006}).

\bibitem[{\citenamefont{Hagley et~al.}(1999)\citenamefont{Hagley, Deng, Kozuma,
  Wen, Helmerson, Rolston, and Phillips}}]{Hagley99b}
\bibinfo{author}{\bibfnamefont{E.~W.} \bibnamefont{Hagley}},
  \bibinfo{author}{\bibfnamefont{L.}~\bibnamefont{Deng}},
  \bibinfo{author}{\bibfnamefont{M.}~\bibnamefont{Kozuma}},
  \bibinfo{author}{\bibfnamefont{J.}~\bibnamefont{Wen}},
  \bibinfo{author}{\bibfnamefont{K.}~\bibnamefont{Helmerson}},
  \bibinfo{author}{\bibfnamefont{S.~L.} \bibnamefont{Rolston}},
  \bibnamefont{and} \bibinfo{author}{\bibfnamefont{W.~D.}
  \bibnamefont{Phillips}}, \bibinfo{journal}{Science}
  \textbf{\bibinfo{volume}{283}}, \bibinfo{pages}{1706} (\bibinfo{year}{1999}).

\bibitem[{\citenamefont{Inouye et~al.}(1999)\citenamefont{Inouye, Pfau, Gupta,
  Chikkatur, G{\"o}rlitz, Pritchard, and Ketterle}}]{Inouye99}
\bibinfo{author}{\bibfnamefont{S.}~\bibnamefont{Inouye}},
  \bibinfo{author}{\bibfnamefont{T.}~\bibnamefont{Pfau}},
  \bibinfo{author}{\bibfnamefont{S.}~\bibnamefont{Gupta}},
  \bibinfo{author}{\bibfnamefont{A.~P.} \bibnamefont{Chikkatur}},
  \bibinfo{author}{\bibfnamefont{A.}~\bibnamefont{G{\"o}rlitz}},
  \bibinfo{author}{\bibfnamefont{D.~E.} \bibnamefont{Pritchard}},
  \bibnamefont{and} \bibinfo{author}{\bibfnamefont{W.}~\bibnamefont{Ketterle}},
  \bibinfo{journal}{Nature} \textbf{\bibinfo{volume}{402}},
  \bibinfo{pages}{641} (\bibinfo{year}{1999}).

\bibitem[{\citenamefont{Kinoshita et~al.}(2006)\citenamefont{Kinoshita, Wenger,
  and Weiss}}]{Kinoshita06}
\bibinfo{author}{\bibfnamefont{T.}~\bibnamefont{Kinoshita}},
  \bibinfo{author}{\bibfnamefont{T.~R.} \bibnamefont{Wenger}},
  \bibnamefont{and} \bibinfo{author}{\bibfnamefont{D.~S.} \bibnamefont{Weiss}},
  \bibinfo{journal}{Nature} \textbf{\bibinfo{volume}{440}},
  \bibinfo{pages}{900} (\bibinfo{year}{2006}).

\bibitem[{\citenamefont{Wu et~al.}(2005{\natexlab{b}})\citenamefont{Wu, Wang,
  Diot, and Prentiss}}]{Wu05b}
\bibinfo{author}{\bibfnamefont{S.}~\bibnamefont{Wu}},
  \bibinfo{author}{\bibfnamefont{Y.}~\bibnamefont{Wang}},
  \bibinfo{author}{\bibfnamefont{Q.}~\bibnamefont{Diot}}, \bibnamefont{and}
  \bibinfo{author}{\bibfnamefont{M.}~\bibnamefont{Prentiss}},
  \bibinfo{journal}{Phys. Rev. A} \textbf{\bibinfo{volume}{71}},
  \bibinfo{pages}{043602} (\bibinfo{year}{2005}{\natexlab{b}}).

\bibitem[{\citenamefont{Reeves et~al.}(2005)\citenamefont{Reeves, Garcia,
  Deissler, Baranowski, Hughes, and Sackett}}]{Reeves05}
\bibinfo{author}{\bibfnamefont{J.~M.} \bibnamefont{Reeves}},
  \bibinfo{author}{\bibfnamefont{O.}~\bibnamefont{Garcia}},
  \bibinfo{author}{\bibfnamefont{B.}~\bibnamefont{Deissler}},
  \bibinfo{author}{\bibfnamefont{K.~L.} \bibnamefont{Baranowski}},
  \bibinfo{author}{\bibfnamefont{K.~J.} \bibnamefont{Hughes}},
  \bibnamefont{and} \bibinfo{author}{\bibfnamefont{C.~A.}
  \bibnamefont{Sackett}}, \bibinfo{journal}{Phys. Rev. A}
  \textbf{\bibinfo{volume}{72}}, \bibinfo{pages}{051605(R)}
  (\bibinfo{year}{2005}).

\bibitem[{\citenamefont{Burke et~al.}(2005)\citenamefont{Burke, Garcia, Hughes,
  Livedalen, and Sackett}}]{Burke05}
\bibinfo{author}{\bibfnamefont{J.~H.~T.} \bibnamefont{Burke}},
  \bibinfo{author}{\bibfnamefont{O.}~\bibnamefont{Garcia}},
  \bibinfo{author}{\bibfnamefont{K.~J.} \bibnamefont{Hughes}},
  \bibinfo{author}{\bibfnamefont{B.}~\bibnamefont{Livedalen}},
  \bibnamefont{and} \bibinfo{author}{\bibfnamefont{C.~A.}
  \bibnamefont{Sackett}}, \bibinfo{journal}{Rev. Sci. Instrum.}
  \textbf{\bibinfo{volume}{76}}, \bibinfo{pages}{116105}
  (\bibinfo{year}{2005}).

\end{thebibliography}

\end{document}